\documentclass[conference]{IEEEtran}
\IEEEoverridecommandlockouts

\usepackage{cite}
\usepackage{amsmath,amssymb,amsfonts}
\usepackage{algorithmic}
\usepackage{graphicx}
\usepackage{textcomp}
\usepackage{xcolor}

\usepackage{tabularx}
\usepackage{array}

\usepackage{censor}

\StopCensoring

\usepackage{booktabs} 

\def\BibTeX{{\rm B\kern-.05em{\sc i\kern-.025em b}\kern-.08em
    T\kern-.1667em\lower.7ex\hbox{E}\kern-.125emX}}

\begin{document}

\title{EEspice: A Modular Circuit Simulation Platform with Parallel Device Model Evaluation via Graph Coloring}

\author{\IEEEauthorblockN{\blackout{Xuanhao Bao}}
\IEEEauthorblockA{\textit{\blackout{School of Engineering}} \\
\textit{\blackout{The University of Edinburgh}}\\
\blackout{Edinburgh, UK} \\
\blackout{X.Bao-10@sms.ed.ac.uk}}
\and
\IEEEauthorblockN{\blackout{Danial Chitnis}}
\IEEEauthorblockA{\textit{\blackout{School of Engineering}} \\
\textit{\blackout{The University of Edinburgh}}\\
\blackout{Edinburgh, UK} \\
\blackout{d.chitnis@ed.ac.uk}}
}

\maketitle

\begin{abstract}
As modern analogue/mixed-signal design increasingly relies on optimization-in-the-loop flows, such as AI and LLM-based sizing agents that repeatedly invoke SPICE—efficient, accurate high-performance simulators have become an indispensable foundation for modern integrated circuit (IC) design. However, the computational cost of evaluating nonlinear models, particularly for BSIM models, remains a significant bottleneck. In standard parallelization approaches, devices such as transistors are easily distributed across processors. The subsequent stamping phase, where each device's contributions are added to the shared system matrix, often creates a bottleneck. Because multiple processor cores compete to update the same matrix elements simultaneously, the system is forced to process tasks one at a time to avoid errors. This paper introduces EEspice, an open-source circuit simulation framework whose modular architecture decouples device model evaluation into independently replaceable kernels, enabling a parallel stamping strategy that overcomes this bottleneck. It partitions MOSFET instances into independent color groups, which can be processed in parallel. Our results show that on a 64-core workstation, the proposed approach achieves up to 45x speedup over single-thread performance when conflicts are low. Our analysis also explores how performance depends on circuit topology.
\end{abstract}


\section{Introduction}
SPICE-like simulators remain the cornerstone of transistor-level verification \cite{nagel1975spice2}, yet AI-driven frameworks such as EEsizer \cite{liu2025eesizer} impose new throughput demands for iterative sizing loops. While high-level abstractions such as real-number models (RNM) and behavioral models exist \cite{balasubramanian2013solutions}, they are primarily used for system-level mixed-signal verification and fast functional abstraction rather than capturing bias-dependent model effects needed for sizing. Sizing tasks require accurate models, such as the BSIM family, to capture bias-dependent nonlinearities. In the transistor-level transient analysis performed by SPICE-like simulators, the circuit equations are solved by Newton--Raphson (NR) iterations \cite{nagel1975spice2}. Each NR iteration evaluates these nonlinear device models at the current operating point, linearizes them, and forms a linearized modified nodal analysis (MNA) \cite{ho1975modified} system \(A x = b\) whose solution updates the unknown node voltages/currents. As optimization-in-the-loop sizing flows repeatedly invoke the simulator, the throughput of this NR loop becomes critical \cite{liu2025eesizer}. Studies show that model evaluation is a dominant cost per NR step, making it a key target for acceleration \cite{5272548}.

Model evaluation in an NR iteration can be separated into two sub-steps. First, each device instance computes terminal currents/charges and the corresponding linearized small-signal parameters about the present operating point. Second, these contributions are stamped, i.e., accumulated into the shared global circuit matrix \(A\) and right-hand-side vector \(b\). The first sub-step is embarrassingly parallel\cite{bayoumi2008massive}\cite{han2013tinyspice}\cite{verley2018xyce}. The second sub-step is harder to scale on a multi-core machine: all instances write into the same \(A\) and \(b\), so if two compute threads update the same matrix entry at the same time, one update can overwrite the other, and the accumulated value becomes invalid due to a data race. A thread lock can prevent this by enforcing one-at-a-time updates for shared entries, but lock contention forces threads to wait and can effectively serialize stamping, limiting the overall speedup. We therefore target lock-free parallelism for the stamping stage and implement it in EEspice. Ngspice \cite{vogt2021Ngspice} is used only as a reference for accuracy and kernel-level profiling because it is fully open-source.

To eliminate thread lock, we schedule stamping so that concurrently processed devices never touch the same matrix elements. We model conflicts with a conflict graph: each vertex represents a device instance, and an edge connects two vertices if they share a non-grounded circuit node. Graph coloring \cite{jones1993parallel} assigns a color to every vertex such that any two connected vertices have different colors. Devices sharing a color are therefore conflict-free and can be processed in parallel. A greedy coloring algorithm \cite{garcia2025greedy} constructs such a coloring by visiting vertices in an order and assigning each vertex the smallest color not used by its already-colored neighbors. Similar coloring-based schedules have been used to avoid write conflicts in other scientific computing tasks \cite{coleman1983estimation}\cite{ccatalyurek2012graph}\cite{krysl2024parallel}.

This paper presents EEspice, an open-source C++ circuit simulation framework whose modular architecture encapsulates device evaluation and stamping as interchangeable kernel modules, allowing different parallelization strategies to be selected without altering the core simulation loop. Exploiting this modularity, EEspice applies graph coloring to the stamping stage: it constructs the FET conflict graph during setup, partitions instances into color groups, and executes the NR loop by processing colors sequentially while evaluating and stamping all devices within each color in parallel, removing stamping data races without locks. Results show coloring restores scalability when conventional threading becomes stamping-limited, and the approach can extend directly to modern devices such as BSIM-CMG \cite{duarte2015bsim}.

\begin{figure}[t!]
    \centering
    \includegraphics[width=0.9\linewidth]{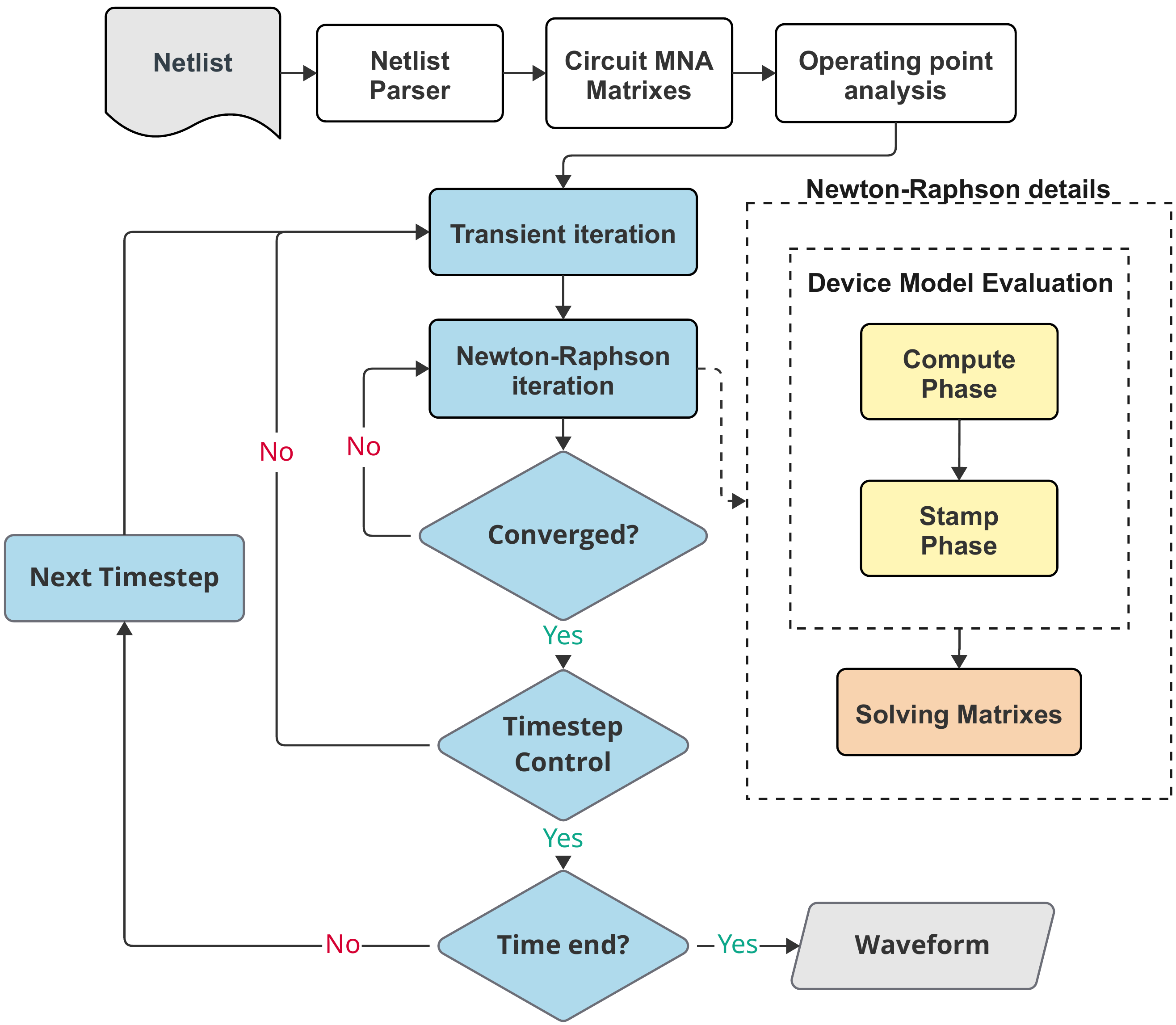}
    \caption{Typical transient simulation flowchart.}
    \label{fig:flowchart}
\end{figure}
\section{Methodology}
\label{lab:Methodology}
\begin{figure}[t!]
    \centering
    \includegraphics[width=1\linewidth]{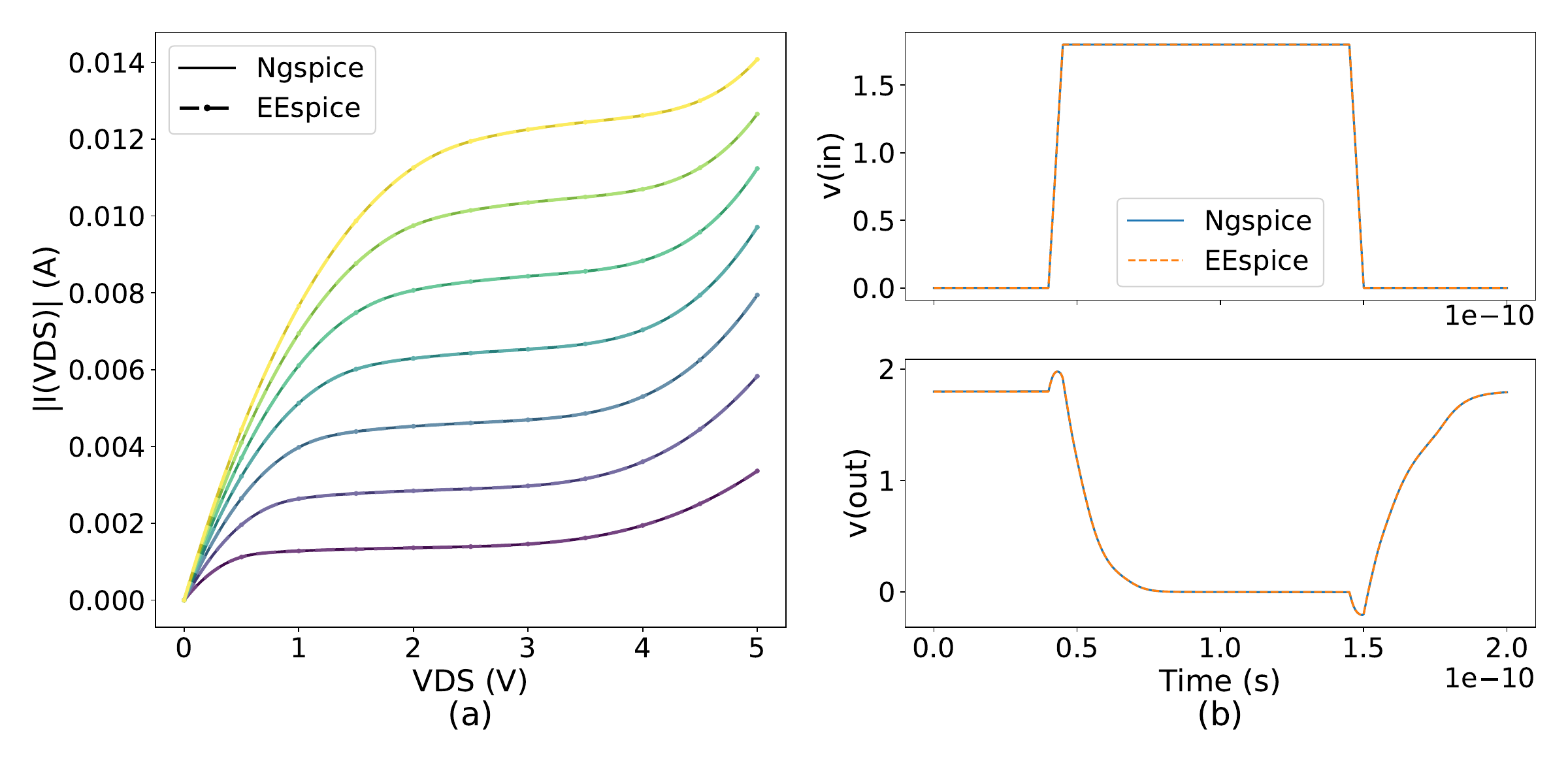}
    \caption{Accuracy comparison versus Ngspice: (a) BSIM4 FET DC $I$--$V$ characteristics overlaid for multiple $V_{GS}$ values, and (b)Inverter transient waveforms ($V(in)$ and $V(out)$) overlaid.}
    \label{fig:combined_plot}
\end{figure}
Fig.~\ref{fig:flowchart} summarizes the transient-analysis loop in EEspice, following the standard SPICE flow. Within each NR iteration, EEspice separates device model evaluation into two phases (Fig.~\ref{fig:flowchart}):
(i) Compute Phase: each device instance evaluates the model at the current bias point, producing linearized contributions---equivalent conductances and current/charge terms for MNA stamping.
(ii) Stamp Phase: per-instance contributions are stamped into the global sparse MNA matrix \(A\) and right-hand-side vector \(b\).

We verify EEspice against Ngspice with identical BSIM4 models and analysis settings. A FET DC sweep (Fig.~\ref{fig:combined_plot}(a)) and a CMOS inverter transient simulation (Fig.~\ref{fig:combined_plot}(b)) produce overlapping traces, confirming performance differences arise solely from the parallelization strategy.

\begin{figure}[t!]
    \centering
    \includegraphics[width=1\linewidth]{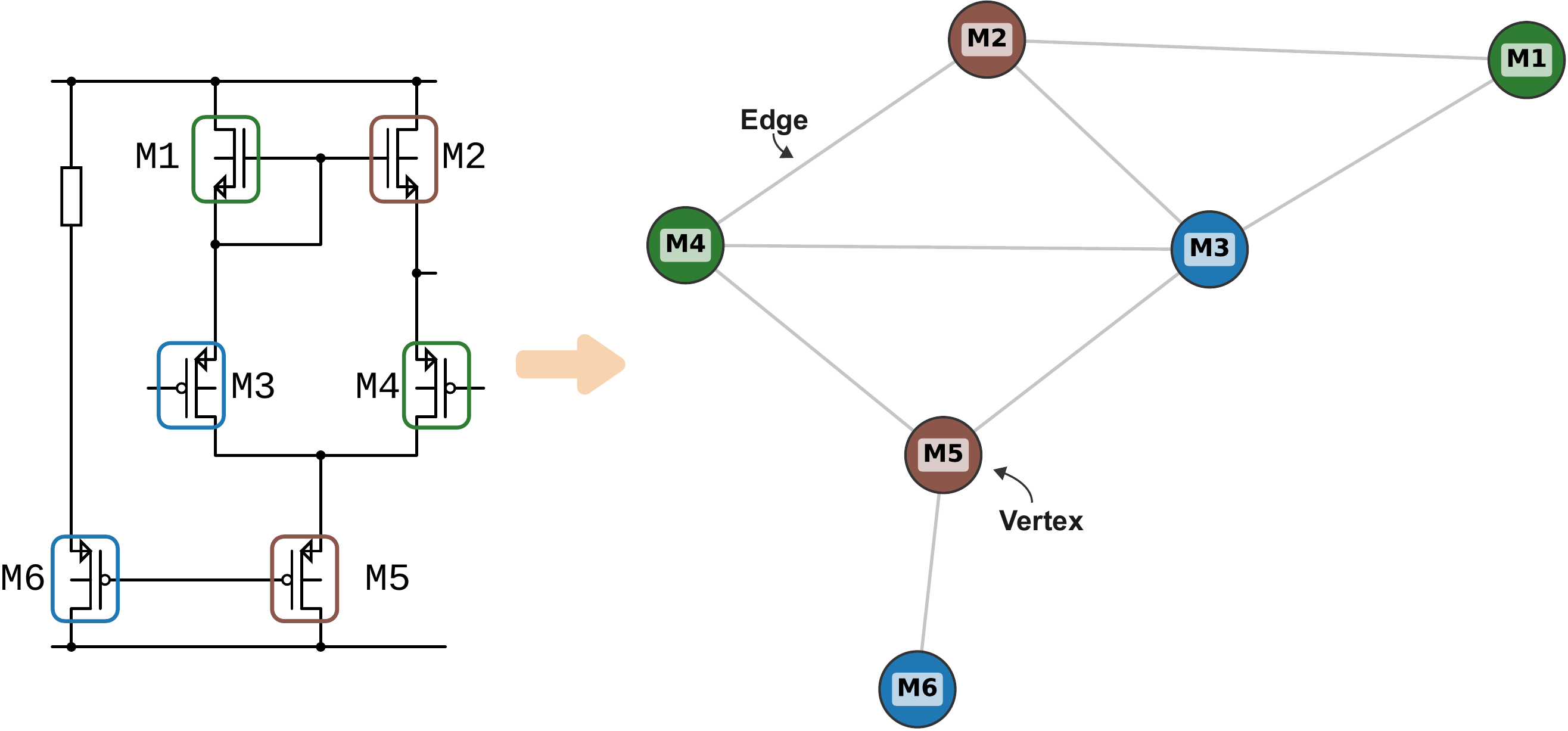}
    \caption{Example of mapping of a 5-transistor Operational Transconductance Amplifier (5T OTA) with a current mirror to a conflict graph for parallel matrix stamping. Each MOSFET instance, such as M1 and M2, is represented as a graph vertex. An edge connects devices that share a common non-ground circuit node, indicating a potential write conflict during stamping. The assigned colors illustrate the partitioning of these devices into independent, conflict-free sets that can be evaluated and stamped concurrently without atomic locks.}
    \label{fig:6tota-coloring}
\end{figure}

To avoid synchronization during stamping, EEspice pre-computes a conflict graph and applies graph coloring. The conflict graph is constructed at circuit setup: each FET instance becomes a vertex. An edge connects two vertices whose FETs share at least one non-ground node, indicating their stamps touch common MNA rows/columns.
Each vertex receives a color such that adjacent vertices differ in color. Same-color devices share no non-ground node and can be stamped concurrently without conflicts. Fig.~\ref{fig:6tota-coloring} illustrates this on a five-transistor operational transconductance amplifier (5T OTA) with a current mirror \cite{yin2023low}.
During the NR loop, colors are processed sequentially. Within each color, all devices are evaluated and stamped in parallel, ensuring correctness without per-entry locks.
\begin{table}[t!]
\centering
\caption{Summary of BSIM4 evaluation-kernel variants implemented in EEspice}
\label{tab:method-summary}
\begin{tabular}{llll}
\hline\hline
Kernel & Compute Phase & Stamp Phase & OpenMP \\
\hline
\emph{loadsingle} & sequential & sequential & No OpenMP \\
\emph{loadomp} & parallel & sequential & \emph{parallel for}\\
\emph{Color} & parallel & parallel & Single \emph{parallel} region \\
\emph{ColorFused} & fused & parallel & Per-color \emph{parallel for} \\
\hline\hline
\end{tabular}
\end{table}
\begin{figure}[t!]
    \centering
    \includegraphics[width=1\linewidth]{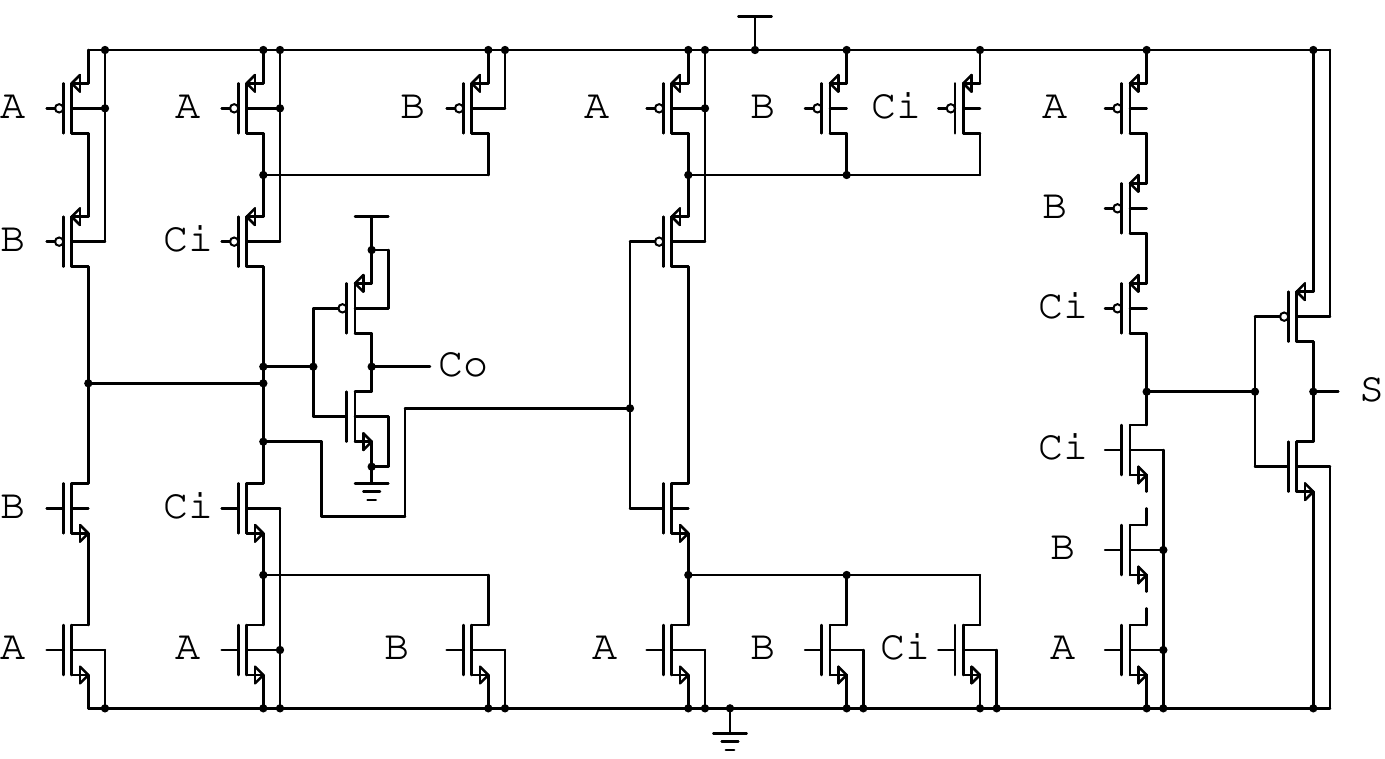}
    \caption{A typical 28T CMOS full adder circuit.}
    \label{fig:schematic}
\end{figure}
All experiments were executed on an AMD Ryzen Threadripper PRO 5995WX (64 cores, 128 hardware threads) with BSIM~4.82. The speedup \(S\) is relative to a single-thread baseline. \(Th\) denotes OpenMP thread count and \(C\) the color count.
OpenMP is a directive-based shared-memory parallel programming model widely used in C/C++, which parallelizes device-instance loops across CPU threads in this work.

\noindent{Benchmark 1: BSIM4 evaluation-kernel.}
This benchmark isolates the BSIM4 evaluation kernel and compares the four variants in Table~\ref{tab:method-summary}. All variants compute identical BSIM4 contributions, differing only in compute/stamp scheduling.
\emph{loadsingle} is the sequential baseline.
\emph{loadomp} parallelizes only the Compute phase, buffering per-device results before sequential stamping.
\emph{Color} parallelizes both phases by a pre-computed color, retaining the same buffer as \emph{loadomp}.
\emph{ColorFused} fuses Compute and Stamp within each color, letting each device stamp immediately after evaluation, eliminating the buffer.
The benchmark sweeps \(Th\) to compare scaling across varying conflict levels (\(C\)).

\noindent{Benchmark 2: 64-bit full-adder transient simulation with topology variants.}
This benchmark evaluates end-to-end transient simulation on a realistic digital netlist: 64 cascaded 1-bit full-adders forming a 64-bit ripple-carry adder (Fig.~\ref{fig:schematic}). Three versions are simulated.
Baseline: the original 64-bit adder netlist.
R14: each 1-bit stage is modified by inserting 14 series resistors of $1 \times 10^{-3}~\Omega$ at randomly selected transistor-to-transistor connections.
R89: each 1-bit stage is modified by inserting 89 series resistors of $1 \times 10^{-3}~\Omega$ at all transistor-to-transistor connections. R14 and R89 reduce stamping conflicts from shared nodes. Inserting small series resistors splits a net into multiple local nodes, each separated by a resistor, so fewer FETs share a node. This reduces edge count in the FET conflict graph and typically lowers \(C\) from the greedy coloring algorithm. All three netlists are simulated in both EEspice and Ngspice to compare numerical behavior and runtime under similar NR iteration counts.

\section{Results}
The results for the two benchmarks are presented in this section.
Unless stated otherwise, all speedups are computed relative to \emph{loadsingle}, the single-threaded EEspice baseline executed with one OpenMP thread.

\begin{figure}[t!]
    \centering
    \includegraphics[width=1\linewidth]{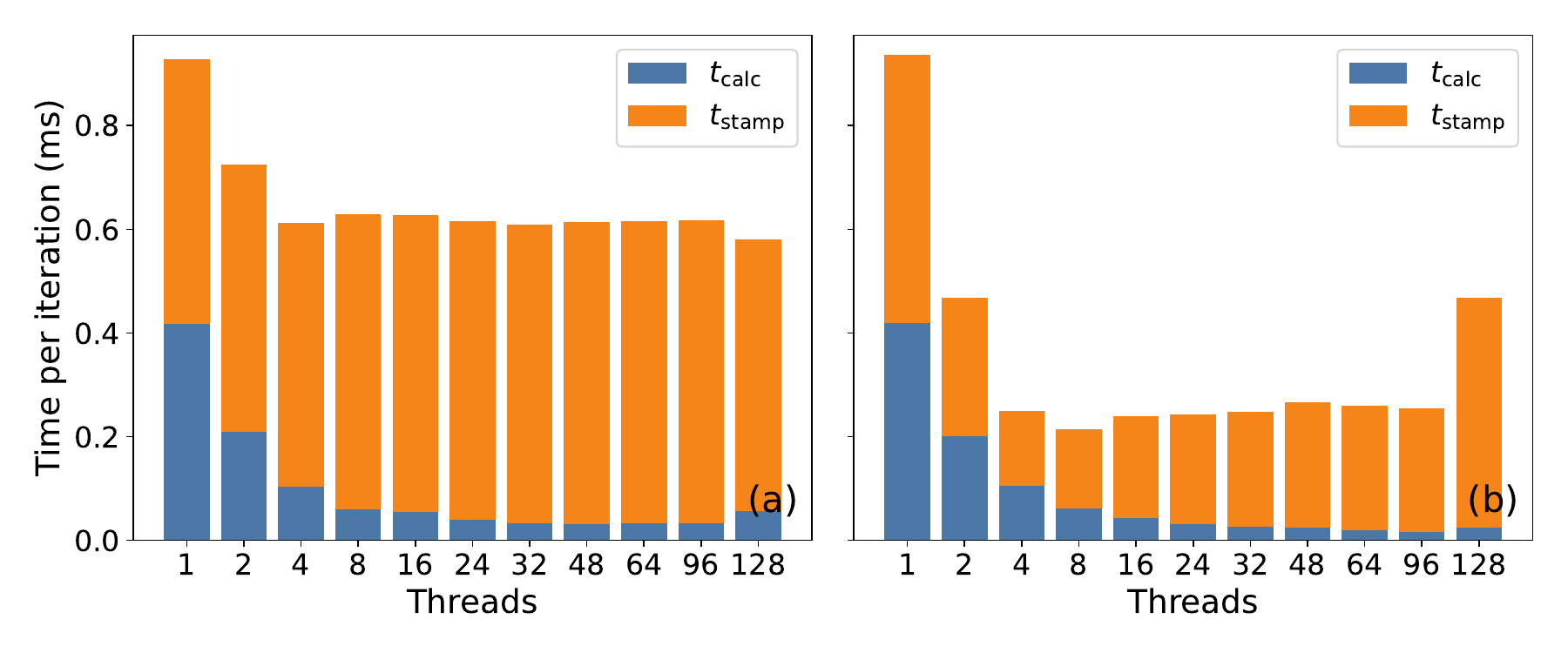}
    \caption{BSIM4 evaluation time breakdown for thread count $N=1000$ under moderate conflict when the number of colors is $C=26$. (a) \emph{loadomp} showing computation time $t_{calc}$ in blue, scaling while stamping time $t_{stamp}$ in orange, remains a serial bottleneck (b) \emph{Color} showing parallelized stamping within color groups, which reduces the total iteration time and prevents the serial bottleneck effect.}
    \label{fig:r1}
\end{figure}

Fig.~\ref{fig:r1} illustrates the BSIM4 kernel cost breakdown. In \emph{loadomp}, only computation ($t_{\mathrm{calc}}$) is parallelized, leaving stamping ($t_{\mathrm{stamp}}$) as a serial bottleneck. As thread count increases, $t_{\mathrm{stamp}}$ remains constant and eventually consumes up to $95\%$ of the total time, capping the maximum speedup near $1.5\times$. In contrast, \emph{Color} parallelizes the stamping phase using conflict-free color groups. This effectively eliminates the serial bottleneck, significantly reducing total NR iteration time and achieving a $4.4\times$ speedup over the single-thread baseline of EEspice.

\begin{figure}[t!]
    \centering
    \includegraphics[width=1\linewidth]{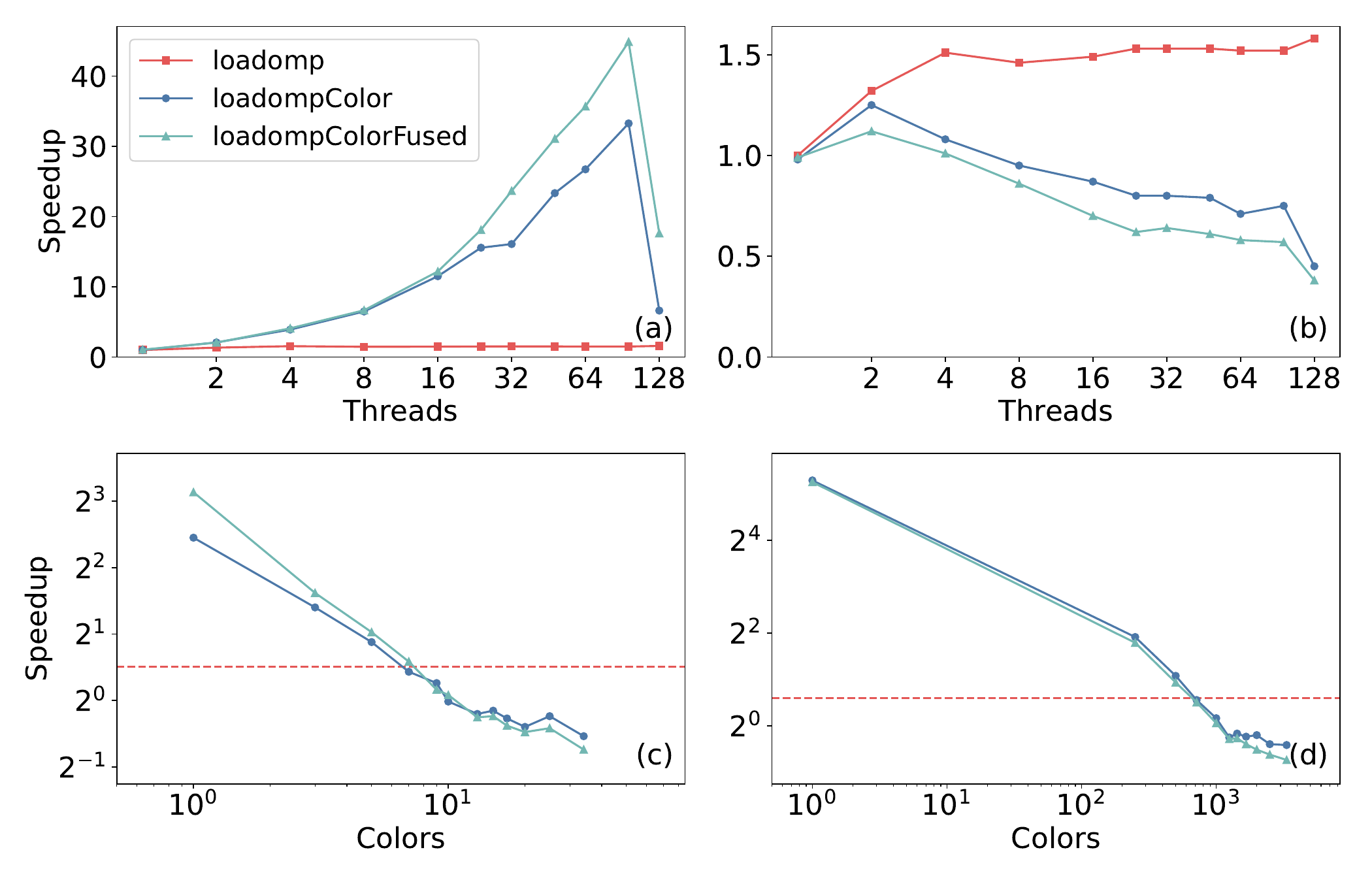}
    \caption{Scaling of coloring-based parallel stamping with thread count and conflict level. The speedup versus OpenMP threads for $N=1000$ in (a) a low-conflict regime $C=1$ and (b) a high-conflict regime $C=334$. The speedup at fixed $Th=64$ versus the number of colors $C$ for (c) $N=100$ instances and (d) $N=10000$ instances on a log axis. \emph{loadomp} is shown as a horizontal reference line. The speedup is measured relative to the single-thread \emph{loadsingle} baseline.}
    \label{fig:r5}
\end{figure}

\begin{table*}[t!]
    \centering
    \caption{Performance Comparison: EEspice vs. Ngspice}
    \label{tab:performance}
    \begin{tabular}{lcccccccc}
    \hline\hline
    Circuits & Transistors & Resistors & Colors & Threads & Device Eval (s) & Matrix Solve (s) & Total Tran Sim (s) & NR It. \\
    \hline
    \multicolumn{9}{l}{\textit{\textbf{EEspice} (ColorFused)}} \\ 
    64-bit Full Adder     & 1792 & 0    & 896 & 64 & 127 & 4.40  & 143 & 50079 \\
    64-bit Full Adder-R14 & 1792 & 896  & 64  & 64 & \textbf{11.6}  & 7.86  & 32.1  & 50083 \\
    64-bit Full Adder-R89 & 1792 & 5696 & \textbf{1}   & 64 & \textbf{3.62}   & 29.6 & 47.2  & 50079 \\
    \hline
    \multicolumn{9}{l}{\textit{\textbf{EEspice} (loadomp)}} \\ 
    64-bit Full Adder     & 1792 & 0    & -- & 64 & 12.1 & 4.19  & 28.6 & 50079 \\
    64-bit Full Adder-R14 & 1792 & 896  & --  & 64 & 12.1  & 7.66  & 32.2  & 50083 \\
    64-bit Full Adder-R89 & 1792 & 5696 & --   & 64 & 12.8   & 29.6 & 56.1  & 50079 \\
    \hline
    \multicolumn{9}{l}{\textit{\textbf{Ngspice}}} \\
    64-bit Full Adder     & 1792 & 0    & -- & 64 & 18.2  & 3.05  & 33.4  & 50034 \\
    64-bit Full Adder-R14 & 1792 & 896  & -- & 64 & 19.5  & 6.20  & 38.2  & 50016 \\
    64-bit Full Adder-R89 & 1792 & 5696 & -- & 64 & 28.7  & 25.9 & 63.4  & 50002 \\
    \hline\hline
    \end{tabular}
\end{table*}

\begin{figure}[t!]
    \centering
    \includegraphics[width=1\linewidth]{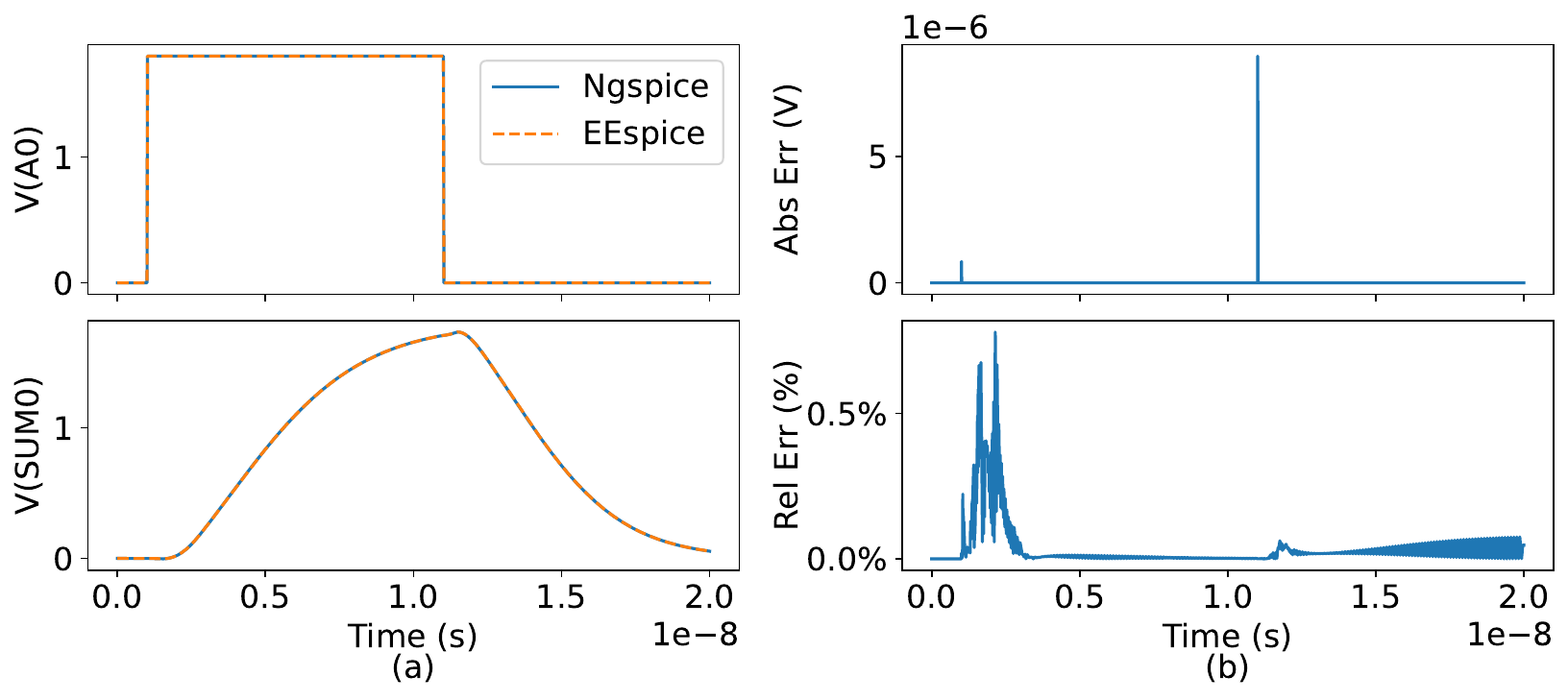}
    \caption{Transient comparison for bit~0 of the 64-bit ripple-carry adder. Node $A0$ is driven by a pulse, and all other inputs are held at $0~\mathrm{V}$. (a) overlays $V(A0)$ and $V(SUM0)$ from Ngspice and EEspice. (b) plots the absolute error in $V(A0)$ and the relative error in $V(SUM0)$ between EEspice and Ngspice.}
    \label{fig:64bit_adder_combined}
\end{figure}

Fig.~\ref{fig:r5} shows how the speedup depends on the thread count $Th$ and the number of colors $C$. In the low-conflict case, $C=1$, Fig.~\ref{fig:r5}(a), all devices share a single color group, so the colored kernels execute the entire workload in one parallel step. \emph{ColorFused} reaches up to $45\times$ speedup, while \emph{loadomp} remains $1.5\times$ due to its serial Stamp phase. Conversely, under high conflict $C=334$, Fig.~\ref{fig:r5}(b), the workload fragments into many small sequential color groups whose per-color synchronization overhead dominates, causing the colored kernels to drop below $1\times$ at high thread counts. Fig.~\ref{fig:r5}(c)--(d) confirms this trend at fixed $Th=64$: speedup decreases with rising $C$, but circuits with more devices (10{,}000 vs.\ 100) sustain useful speedup to a larger $C$ because each color group retains enough work to offset overhead.

Fig.~\ref{fig:64bit_adder_combined} compares EEspice and Ngspice on a 64-bit ripple-carry adder transient waveforms. Input $A0$ (bit~0) is pulsed while all other inputs are held at $0~\mathrm{V}$. The simulated waveforms for $V(A0)$ and $V(SUM0)$ overlap between the two simulators in Fig.~\ref{fig:64bit_adder_combined}a. The relative error below $1\%$ over the time window in Fig.~\ref{fig:64bit_adder_combined}b.

Table~\ref{tab:performance} summarizes the simulation performance across the three 64-bit adder netlists, focusing on device evaluation time. 
In the unmodified adder, heavy node sharing yields 896 color groups. This makes the \emph{ColorFused} evaluation approximately 127~s, which is relatively inefficient, whereas EEspice \emph{loadomp} evaluates much faster, finishing at 12.1~s and Ngspice 18.2~s.
Modifying the topology by adding resistors reduces shared nodes and color counts, allowing \emph{ColorFused} to perform optimally. For the R14 netlist with $C=64$, its evaluation time drops to 11.6~s, outperforming both Ngspice and \emph{loadomp}. For the R89 netlist with $C=1$, evaluation time further minimizes to 3.62~s. 
However, the inserted resistors increase the sparsity of the linear system. Since both EEspice and Ngspice use KLU as their sparse linear solver, the bottleneck shifts to the matrix solve stage, in EEspice to 29.6~s, thereby increasing the total simulation time.

\section{Discussion}
Benchmark~1 demonstrates that device evaluation acceleration is bounded by the stamping phase. While \emph{loadomp} saturates at $1.5\times$ due to sequential stamping (Fig.~\ref{fig:r1}(a)), coloring-based kernels enable lock-free parallel updates. Efficiency here depends on the color count $C$: low $C$ allows \emph{ColorFused} to scale up to $45\times$, whereas high $C$ introduces synchronization overhead that diminishes gains (Fig.~\ref{fig:r5}).

Benchmark~2 highlights the trade-off between evaluation efficiency and matrix complexity. In the 64-bit adder, high node sharing with $C=896$, makes \emph{loadomp} kernel faster due to color overhead. Conversely, introducing series resistors reduces $C$, enabling fast fused evaluation to reach $3.62$~s for R89, but the resulting larger MNA system shifts the primary bottleneck to the sparse solver at$29.6$~s, resulting in a larger total simulation time.

Crucially, these results show that full end-to-end speedup also depends on solver performance. Once device evaluation is minimized, the solver's computation time dominates. Consequently, accelerating the sparse linear solver is essential to realizing a full transient-loop speedup. While high-performance solvers such as Intel's oneMKL PARDISO and NVIDIA's cuDSS exist, they must be adapted for SPICE-specific matrix structures and factorization patterns. Finally, as post-layout parasitics naturally reduce node sharing, this coloring approach remains highly viable for realistic IC design flows and other models such as BSIM-CMG.

\section{Conclusion}
This work presents EEspice, a simulator whose modular kernel architecture enables graph coloring-based parallel model evaluation, removing matrix stamping conflicts without locks in SPICE transient analysis. By grouping FET instances into conflict-free colors and fusing evaluation with stamping, EEspice achieves up to 45x speedup on 64 cores while matching Ngspice accuracy. Future work should accelerate sparse solvers to reduce the total transient simulation time. The source code for this project is available on \blackout{https://github.com/eelab-dev}

\bibliographystyle{IEEEtran} 
\bibliography{reference}

\end{document}